\documentclass[aps,prl,showpacs,twocolumn,floatfix]{revtex4-1}

\usepackage{amsmath,amssymb,epsfig,amsbsy,xspace,empheq}
\usepackage{color}
\usepackage{subfigure}
\usepackage{multirow}
\usepackage{hyperref}

\newcommand{\be}{\begin{equation}}
\newcommand{\ee}{\end{equation}}

\newcommand{\ZZ}{\mathbb{Z}_}

\begin{document}

\title{Naturally light scalar particles: a generic and simple mechanism}

\author{F. L\'eonard, B. Delamotte}
\affiliation{CNRS, Sorbonne Universit\'e,  Laboratoire de Physique Th\'eorique de la Mati\`ere Condens\'ee, LPTMC, 
F-75005 Paris, France}
\author{Nicol\'as Wschebor}
\affiliation{Instituto de F\'{\i}sica, Facultad de Ingenier\'{\i}a, 
Universidad de la Rep\'ublica, 11000 Montevideo, Uruguay}

\begin{abstract}

The hierarchy problem in the Standard Model is usually understood as both a technical problem
of stability of the calculation of the quantum corrections to the masses of the Higgs sector
and of the unnatural difference between the Planck and  gauge breaking scales. Leaving aside
the gauge sector, we implement on a purely scalar model a mechanism for generating naturally light scalar
particles where both of these issues are solved.
In this model, on top of terms invariant under a continuous symmetry, a highly non-renormalizable 
term is added to the action that explicitly breaks this symmetry down to a discrete one. 
In the spontaneously broken phase, the mass of the pseudo-Goldstone is then driven by quantum fluctuations 
to values that are non-vanishing but that are generically, that is,  without fine-tuning, orders of magnitude smaller than the UV scale.

\end{abstract}

\maketitle

Solving the  gauge hierarchy problem of the Standard Model (SM) 
has triggered an enormous amount of works these last fourty years (see, for example,
\cite{Dine:2015xga}). 
It has been, in particular, one of the important motivations for studying  
supersymmetric extensions of the SM (for a classical review on the topic, see \cite{Djouadi:2005gj}). Other hypotheses for solving 
this problem have been put forward and intensively studied. Among them,
the most popular have been
the existence of extra dimensions of space-time \cite{ArkaniHamed:1998rs,Randall:1999ee}
or of new strong  
gauge interactions, such as many variants of the Technicolor model \cite{PhysRevD.13.974,PhysRevD.20.2619}. 
Up to now, all these approaches 
show two rather severe drawbacks: They are not supported by
experimental data \cite{Patrignani:2016xqp} and they are rather drastic and sometimes {\it ad hoc } 
modifications of the original SM which is a very accurate description
of the low energy world. 

All the mechanisms mentioned above that aim at solving the hierarchy problem,
take for granted that the SM cannot be made natural, that is, 
cannot avoid fine-tuning without drastic modifications. This belief 
is largely based on an
analysis of the renormalization of the mass of the Higgs 
that shows that it is corrected at the quantum level by 
terms of the order of the ultraviolet (UV) cutoff of the theory, usually 
taken as the Planck or grand unification scale. That the Higgs mass 
be small compared to the UV cutoff thus requires a cancellation 
between large terms -- the bare mass and the quantum corrections --
that makes unnatural the SM \cite{tHooft:1979rat}. For its part, 
supersymmetry  decreases the sensitivity of the mass of the 
Higgs to the UV scale but does not explain {\it per se} the huge difference 
between the Planck and electroweak (EW) scales.\footnote{There are, however,
models based on supersymmetry that explain also the large hierarchy of scales
\cite{Witten:1981nf}.}
Of course, the scale 
at which physics beyond the SM takes place could be well below the Planck scale. 
However, the precision tests of the SM rule out that it be of the order of the Higgs
mass and already require a good deal of fine-tuning (see, for example, \cite{PhysRevD.96.095036}).

The gauge hierarchy problem can be explained from a simple 
analogy with statistical mechanics. A statistical system such as the Ising 
model can be close to criticality, that is, its correlation length 
can be large compared to the microscopic scale of the model (e.g. 
the lattice spacing), only when a fine-tuning of some external 
parameter (e.g. a temperature) has been performed. The deep analogy 
between particle physics and statistical mechanics comes from the 
fact that the role played by the correlation length 
in the latter is exactly the same as the one played by the Compton 
wavelength (the inverse of the mass) in the former. The same holds true
for the roles played by (the inverse of) the Planck scale and the lattice
spacing. The fine-tuning necessary in both cases to get 
correlation/Compton lengths large compared to the UV length scales
is therefore exactly the same from a formal viewpoint. The difference, of course, is
that it can be performed at will in condensed matter systems while 
it is not so in particle physics where the parameters of the SM are 
what they are and thus cannot be tuned. The scalar sector of the 
SM appears therefore as an almost critical model where the fine-tuning remains 
unexplained at least if the ultraviolet scale is  large 
compared to the EW scale and if the particle content is not supplemented
by some extra degrees of freedom and/or symmetries. 

Given the previous discussion, we focus here on the purely scalar sector leaving
for future studies the coupling to the gauge and matter sectors. 
Accordingly, the problem we solve in this Letter is: 
 How can we generate small masses in a purely scalar model without fine-tuning at the UV scale?
This is achieved by slightly modifying the standard $\varphi^4$
scalar model in a way
that drastically changes the quantum (or statistical) corrections to the masses  in the spontaneously broken phase.
For the sake of simplicity, we present the mechanism at work on a simple, two-scalar-field model 
that applies both in high 
energy physics and for condensed matter systems.

An obvious candidate for generating small masses without fine-tuning
is to use spontaneous symmetry breaking of a continuous symmetry that 
produce Goldstone modes. In this case, the problem is twofold. First, the masses
are not only small, they vanish. Second, to get nonvanishing and small masses, it seems necessary
to add small terms in the action that break explicitly the continuous symmetry 
so as to avoid strictly massless particles. In such a case, the scalar particle is called
a pseudo-Goldstone \footnote{Models where the Higgs particle is interpreted as a pseudo-Goldstone
has a long history, see for example \cite{Georgi:2007zza} and references therein.}. However, whatever small these terms are, they are expected to
generate large quantum corrections to the masses. As a result, a fine-tuning 
in the bare action is necessary to get rid of these large quantum corrections and all the good properties
of the initial Goldstone modes are lost (see, for example, chapter 19 of \cite{Weinberg:1996kr}).

An important point for our approach is that the SM is an
effective field theory valid only below a UV scale $\Lambda$.
There is therefore no reason why its bare action should involve only renormalizable terms
(see, for example, chapter 12 of \cite{Weinberg:1995mt}).
Our idea is thus to use the Goldstone mechanism as a starting point 
 but to explicitly break the initial 
continuous symmetry by adding to the action $S$ a
(possibly highly) nonrenormalizable term $\Delta S_{\rm disc.}$, 
instead of the typical breaking by renormalizable  terms (the index disc. 
in $\Delta S_{\rm disc.}$ is here for future convenience). At first sight, 
$\Delta S_{\rm disc.}$ could seem to play no role because a nonrenormalizable
term contributes to renormalize the other couplings in the UV but becomes 
immaterial in the infrared limit \cite{Polchinski:1983gv}.
This is why nonrenormalizable terms are in general 
not considered. However, it is clear that the above argument cannot be 
fully correct because even when it is nonrenormalizable, a well-chosen term 
can explicitly break a continuous symmetry and thus prevent the existence of
strictly massless Goldstone bosons. 
Instead, the mass of the boson is proportional to the coupling constant
in $\Delta S_{\rm disc.}$. 

As shown below, when the model is not too far from its massless limit, that is, from criticality
in the language of statistical mechanics, the
coupling constant in $\Delta S_{\rm disc.}$ falls off very rapidly with the renormalization group (RG) scale, that is, not logarithmically 
but as a power law, and so does the mass of 
the pseudo-Goldstone boson. As a result, instead of being naturally of the order
of $\Lambda$, the mass of the scalar boson is naturally driven by the RG flow to very
small values compared to $\Lambda$: The smallness of the scalar mass is not only stabilized with respect to
the quantum corrections, it is explained by a spontaneous symmetry breaking mechanism. 

Let us now show in detail how this mechanism works on the simplest -- and 
relevant in statistical mechanics -- example of the $SO(2)$ group explicitly broken down to $ \ZZ q$. 
We consider two scalar fields $\boldsymbol{\varphi}=(\varphi_{1},\varphi_{2})$ and the $SO(2)$-invariant euclidean action:
\begin{equation}
 S=\int d^4 x\left[\frac 1 2 (\partial_\mu \boldsymbol{\varphi})^2 +\frac 1 2 m_0^2 \boldsymbol{\varphi}^2
 + \frac{u_0}{8}(\boldsymbol{\varphi}^2)^2\right] .
\end{equation}
When the renormalized square mass is negative, the $SO(2)$ symmetry is spontaneously broken and the transverse mode is the Goldstone mode. 
To this action we add a term $\Delta S_{\rm disc.}$ proportional to $\sigma$, a polynomial of degree $q$, 
that breaks explicitly $SO(2)$ down to $ \ZZ q$. 
For instance, for $q=6$, $\sigma=(\varphi_1 - \varphi_2)^2(\varphi_1^2+4\varphi_1\varphi_2+\varphi_2^2)^2/8 $. 
It is easy to show that all polynomial terms contributing to the potential that are $ \ZZ 6$-invariant 
are built out of $\rho=\frac 1 2 \boldsymbol{\varphi}^2$ and $\sigma$. Of course, since the 
$ \ZZ q$-invariant action involves terms that are not renormalizable for $q>4$, the model should be 
considered as an effective field theory valid below a UV scale $\Lambda$.

The RG flows presented below have been obtained using Wilson's RG \cite{PhysRevB.4.3174} that has the advantages 
of allowing for nonperturbative approximations, of dealing easily with many coupling 
constants (be they renormalizable or not) and of avoiding all the infrared problems of 
the usual unregularized perturbation theory in the presence of Goldstone modes as well 
as all the UV problems related to the presence of nonrenormalizable terms. However, 
for the sake of simplicity, we also give in the following  the one-loop flows obtained by expanding 
the Wilsonian, nonperturbative RG (NPRG) equations at lowest order in the couplings.

The NPRG is based on Wilson's idea of integrating
fluctuations step by step. In its modern version, it
is implemented on the one-particle irreducible  generating functional
$\Gamma$ (the effective action) which is the Gibbs free energy in statistical mechanics
(for original references, see \cite{Wetterich199390,Ellwanger:1993kk,Morris:1993qb}; 
for a pedagogical introduction, see \cite{Delamotte:2007pf} and 
\cite{canet03,canet05,kloss14,delamotte04,canet04,tisser08,canet16,canet06} for applications). 
It amounts to regularizing the low-energy modes, that is, the modes of $\boldsymbol\varphi(q)$ with
wavenumbers $|q| < k$, by giving them a mass
while keeping unchanged the other ones, that is, $\boldsymbol\varphi(|q| > k)$. A
one-parameter family of models indexed by a scale $k$ is
thus defined by 
\begin{equation}
\label{Z}
 {\cal Z}_k[J_i]= \int D\varphi_i \exp( -S[\boldsymbol\varphi]-\Delta S_k[\boldsymbol\varphi]+ \boldsymbol{J}\cdot\boldsymbol\varphi)
\end{equation}
with  $\Delta S_k[\varphi]=\frac 1 2\int_q R_k(q^2) \varphi_i(q)\varphi_i(-q)$ and,
for instance \cite{Litim:2002cf}, 
$R_k(q^2)=Z_k(k^2-q^2)\theta(k^2-q^2)$ with $\theta$
the step function and $Z_k$ the field renormalization constant, to be defined later, and 
with $\boldsymbol{J}\cdot\boldsymbol\varphi=\int_x J_i(x) \varphi_i(x)$. 
The classical field is computed with  ${\cal Z}_k$: $\phi_i(x)=\langle \varphi_i(x)\rangle$.
The $k$-dependent  effective action  $\Gamma_k[\phi_i]$
is defined as usual as the Legendre transform of ${\cal W}_k[J_i]=\log  {\cal Z}_k[J_i]$:
\begin{equation}
\label{legendre}
 \Gamma_k[\boldsymbol\phi]+  {\cal W}_k[\boldsymbol{J}]= \boldsymbol{J}\cdot\boldsymbol\phi-\frac 1 2 \int_q R_k(q^2) \phi_i(q)\phi_i(-q)
 \end{equation}
which, for convenience, is  slightly modified by the inclusion of the last term. 
When $k$ is large, all fluctuations in ${\cal Z}_k$
are frozen by the $R_k$ term 
and the classical approximation becomes exact for $k\simeq\Lambda$. With the definition above, Eq. (\ref{legendre}), 
it is then possible to show that $\Gamma_{k=\Lambda}[\phi_i]\simeq S[\phi_i]$ if $\Lambda$ is  very 
large compared to all other momentum scales \cite{Wetterich199390,Delamotte:2007pf}. 
Since $R_{k=0}(q^2)\equiv 0$,  $\Gamma_{k=0}[\phi]$ 
is the effective action  of the original model, that is, the quantity we want to compute. 
Thus, $\Gamma_k[\phi_i]$ interpolates between the bare action when $k=\Lambda$ and the 
effective action of the model under study when $k=0$. 

The exact RG flow equation of $\Gamma_k$ reads \cite{Wetterich199390}:
\begin{equation}
\label{flow}
\partial_t\Gamma_k[\boldsymbol\phi]=\frac 1 2 {\rm Tr} \left[\partial_t R_k(q^2) (\Gamma_k^{(2)}[q;\boldsymbol\phi]+R_k(q))^{-1}\right]
\end{equation}
where $t=\log(k/\Lambda)$, the trace stands for an integral over $q$ and a trace over group indices and 
$\Gamma_k^{(2)}[q;\boldsymbol\phi]$ is the matrix of the Fourier transforms of the second functional derivatives of 
$\Gamma_k[\boldsymbol\phi]$ with respect to $\phi_i(x)$ and $\phi_j(y)$. Starting with the initial condition 
$\Gamma_{k=\Lambda}[\phi_i]= S[\phi_i]$,  this flow equation leads at $k=0$ to an exact solution of the problem. 
For the systems we are interested in, it is impossible to solve Eq. (\ref{flow}) exactly and we 
therefore have recourse to approximations. 

At tree level and in the broken phase, the field acquires a nonvanishing 
vacuum expectation value (vev) $\sqrt{2\kappa_0}$ in one of the $q$ minima of the potential, that is, 
$\rho_{\vert_{\rm min}}=\kappa_0$ and $\sigma_{\vert_{\rm min}}=0$. The spectrum of the model consists 
of two massive modes. One is the pseudo-Goldstone boson of mass $m_T$ and the other one corresponds to the longitudinal mode
of mass $m_L$. In the $ \ZZ 6$ case where we parameterize the tree-level potential in the following way:
$U(\rho,\sigma)=\frac{u_0}{2}(\rho-\kappa_0)^2+ \lambda_{6,0} \sigma$, we find 
$m_{T,0}^2=18 \lambda_{6,0} \kappa_0^2$ and $m_{L,0}^2=2 u_0\kappa_0$ (here, again, $\rho=\boldsymbol\phi^2/2$). 
Notice that $m_{T,0}=0$ when $\lambda_{6,0}=0$, as expected.

\begin{figure}
 \centering\includegraphics[scale=1]{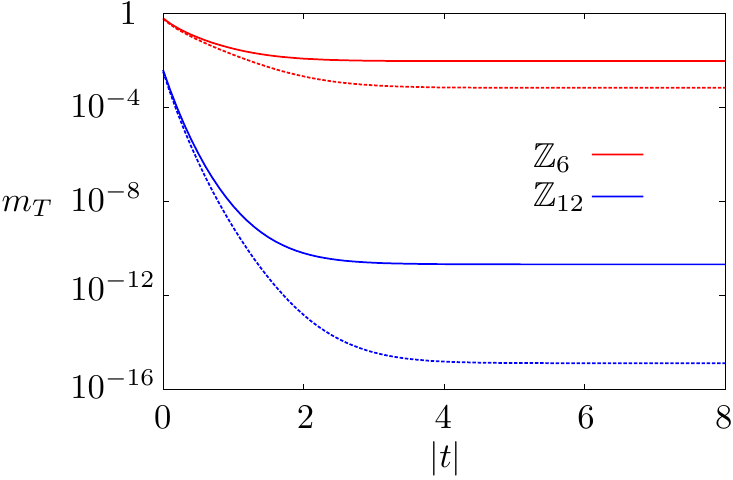} 
 \caption{Flows of the  mass of the transverse mode for the $ \ZZ 6$ and $ \ZZ {12}$ models with two initial conditions. In
 both cases, the upper curves are obtained for a bare mass which is $10\%$ below  its critical value corresponding to the 
 massless theory and the lower ones to a tuning of order $1\%$ below  the bare critical value.}
 \label{Z6Z12}
\end{figure}

We now discuss the one-loop RG flow equations. In the model above, which is regularized in the infrared by the $R_k(q)$ term, the 
couplings and the masses become $k$-dependent and their one-loop flow can be obtained from (\ref{flow}) 
by making the replacement of the full propagator 
$(\Gamma_k^{(2)}[q;\boldsymbol\phi]+R_k(q))^{-1}$ by  $(S_k^{(2)}[q;\boldsymbol\phi]+R_k(q))^{-1}$ 
where $S_k$ is identical to $S$ up to the replacement of the bare couplings by $k$-dependent ones: 
$u_0\to u(k),\ \kappa_0\to\kappa(k),\ \lambda_{6,0}\to\lambda_{6}(k),\ m_{T,0}\to m_{T}(k),\ m_{L,0}\to m_{L}(k)$. 
The corresponding flow equations read \cite{Leonard:2015wyg}
for the function $R_k(q)$ given below Eq.~(\ref{Z}):
\begin{subequations}
 \label{flowLPA}
 \begin{empheq}{align}
 \displaystyle{\partial_t \kappa} = & \displaystyle{\alpha \left(\left[ 1 + 4\frac{m_T^2}{m_L^2 } \right]k^2 I_2( m_{T}^2)
 + 3 k^2 I_2( m_{L}^2)\right)}\\
\displaystyle{\partial_t u} = & \displaystyle{\alpha \left( - 36 \lambda_6 k^2 I_2( m_{T}^2)+ 18 {u}^2 I_3( m_{L}^2)\right.} \nonumber\\
                              & \displaystyle{\left.\ \ \ \ \ \  + 2(  u + 36 \kappa \lambda_6 )^2 I_3( m_{T}^2)\right)}\\
\displaystyle{\partial_t \lambda_6 }=
     &\displaystyle{30\alpha  \lambda_6 k^2(u + 6 \kappa \lambda_6 )\frac{ I_2( m_{T}^2) - I_2( m_{L}^2)}{ m_{L}^2- m_{T}^2}}     
\end{empheq}
\end{subequations}
with  $I_n(m^2)=  (1 + m^2/k^2)^{-n}$ and $\alpha^{-1}=32\pi^2$. Notice that the flow of $\lambda_6$
vanishes when $\lambda_6=0$ which is expected since the $ \ZZ 6$ symmetry is enlarged to SO(2) in this case. 
This is what changes drastically the flow
of this coupling compared to the flow of the coupling of same degree in front of the SO(2)-invariant term: $(\rho-\kappa)^3$ 
(not included here for simplicity).

For $k$ of the order of the masses, the flows depend on the precise shape of the 
function $R_k$ as expected. However, when $k$ is very large compared to the masses, $k\gg m_{T,L}$, the flow 
is particularly simple and $\partial_t u$ and $\partial_t \lambda_6$ become scheme-independent, 
that is, is independent of the choice of regulator $R_k(q)$. In this case, it reads:
\begin{subequations}
 \label{flow1}
 \begin{empheq}{align}
 \displaystyle{  \partial_t\kappa} = &\; \displaystyle{4\alpha\,\frac{m_L^2+m_T^2}{m_L^2}k^2}\\
 \displaystyle{\partial_t  u}      = & \;\displaystyle{2\alpha\left( 9 u^2-18\lambda_6 k^2+(u+36\kappa\lambda_6)^2\right)}\\					 
 \displaystyle{  \partial_t \lambda_6 }  = & \;\displaystyle{60\alpha\lambda_6 (u+6\kappa\lambda_6) .}                                            
\end{empheq}
\end{subequations}
A second important property
of the flow is that whenever $k$ becomes smaller than a given mass, the corresponding degree of 
freedom decouples from the flow since its fluctuations become negligible at momentum scales
smaller than its mass.
In practice, Eqs. (\ref{flowLPA}) are convenient because they automatically take into account these two aspects of the flow.

Given that we are {\it a priori} interested in couplings that can be large, the accuracy 
of the one-loop results derived from Eqs. (\ref{flowLPA}) could be questionable. We have checked that they are indeed robust at least at
a qualitative and even semi-quantitative level by considering a celebrated and nonperturbative approximation 
of the exact flow, Eq. (\ref{flow}), called the
local potential approximation prime (LPA') \cite{Berges2002223,Canet:2002gs}. It consists in 
substituting in Eq. (\ref{flow}) an {\it ansatz} for $\Gamma_k$ under the form:
\begin{equation}
 \Gamma_k[\boldsymbol\phi]\to \Gamma_k^{\rm LPA}[\boldsymbol\phi]=
 \int_x\,\left[\frac 1 2 Z_k(\partial_\mu\boldsymbol\phi)^2+ U_k(\rho,\sigma)\right]
\end{equation}
where $Z_k$ is the field renormalization and $U_k(\rho,\sigma)$ a general function of $\rho$ and $\sigma$. 
The LPA' above misses of course all derivative terms of orders higher than two. Its accuracy stems
from the fact that we are only interested  in the flows of the couplings of the potential that are only weakly 
impacted by neglecting higher derivative terms.  This approximation has been shown to work extremely well for $ \ZZ q$-symmetric models
in $d=3$ where the couplings are large and the field renormalization small
but not fully negligible \cite{Leonard:2015wyg}.
It should even work better in dimension four.

When the potential $U_k(\rho,\sigma)$ is truncated by including only the couplings $\kappa(k), u(k),\lambda_{6}(k)$,
the LPA' flow boils down to Eqs. (\ref{flowLPA}). We have checked that the flow of the masses converges with the order
of the expansion of $U_k(\rho,\sigma)$ around the vev of the field when 
including more and more powers of $\rho-\kappa$ and $\sigma$.
We have also checked that keeping the field renormalization
$Z_k$ or approximating it by $Z_k=1$ for all $k$ changes only slightly our results 
and does not spoil qualitatively our conclusions. Our analysis shows that keeping 
only the couplings $\kappa(k), u(k),\lambda_{6}(k)$ already gives the correct general picture of the flow.
The flow equation for the potential $U_k(\rho,\sigma)$ in the $ \ZZ 6$ case is given in the 
Supplemental Material (and for the $ \ZZ {12}$ case on request).

We provide in Fig. (\ref{Z6Z12}) the flows of the mass $m_T$ of the transverse mode in the $ \ZZ 6$ and $ \ZZ {12}$ cases. 
They have been obtained within the LPA' by taking 
a natural  bare action as initial condition of the RG flow, that is, with $u(k=\Lambda)=u_0=\alpha^{-1}$,
$\lambda_{6}(k=\Lambda)=\lambda_{6,0}=\alpha^{-2}\Lambda^2$ and  the couplings of the higher order terms equal to 0. 
The bare mass $m^2_L(k=\Lambda)=2 \kappa_0 u_0$ is chosen either $10\%$ or $1\%$ below the critical 
value that makes the model massless \footnote{As said before, a one-loop approximation gives 
qualitatively the same results.}.
It is clear from Fig. (\ref{Z6Z12}) that the mass decreases dramatically in the first RG 
steps and then saturates to a value which is much lower 
for $ \ZZ {12}$ than for $ \ZZ 6$. This result is easy to understand from the remark that 
the more irrelevant $\sigma$, the faster
the decrease of its coupling constant. This is due to the fact that when $q$ grows, the symmetry
$ \ZZ {q}$ of the model becomes closer to the full continuous SO(2) symmetry. It is also 
clear that the closer to the massless case, the smaller the final value of the mass since its decrease
takes place on a longer RG ``time'' $\vert t\vert$. In all cases, we find that it is easy to obtain a mass $m_T$
of the transverse mode orders of magnitude smaller than the UV scale
 and it can easily be $10^{10}$ times smaller for $ \ZZ {12}$.

To conclude, let us remark that the idea that the smallness of some physical observables 
can be associated with some non-renormalizable terms is not new. For example, this is the 
case for the well-known mechanism that generates small masses for neutrinos from operators of
dimension five \cite{Weinberg:1979sa}. This is, of course, also the case for gravity which 
is supposed to be very small because its interaction terms are
not renormalizable \cite{Weinberg:1995mt}. Let us point out, however, that these effects 
are present already at tree level. The mechanism proposed above for generating small masses 
for scalar particles is, in this respect, different because it is associated with the fluctuations 
of the fields and is absent at tree level. More precisely, we have shown that it is easy to 
generate small masses from  spontaneous symmetry breaking if some non-renormalizable terms break a 
continuous group down to a discrete one. This can be done without fine-tuning of the parameters of the bare action.
In a sense, the fluctuations are no longer a problem but a solution to the problem of the smallness of the
scalar mass.

It is interesting to point out that the very same mechanism has been studied in 
statistical mechanics and is responsible for a striking phenomenon: The critical exponents 
associated with a second order phase transition are not the same in the symmetric and broken 
phases when non-renormalizable discrete anisotropies are present \cite{Leonard:2015wyg}. 
In this context such terms are called ``dangerously irrelevant'' \cite{PhysRevB.13.2222}.

The present work shed new light on the Standard Model hierarchy problem 
because it is usually admitted that the main difficulty is to generate small masses for 
scalar particles without fine-tuning. Even if the present mechanism shows how to 
generate such small masses, it is not obvious how to employ this idea  to 
generate small masses in the gauge sector.  Finally, we notice that the study 
of the interplay between apparent fine-tunings in various physical theories and fluctuations associated with 
non-renormalizable terms has not been studied in depth in the literature. We believe
that this direction of research could be very fruitful as suggested by the model studied above.

N.W. acknowledges Ecole Polytechnique, Palaiseau,  where part of this work has been done
and also supports from Grant No. 412 FQ 293 of the CSIC (UdelaR) commission
and Programa de Desarrollo de las Ciencias B\'asicas (PEDECIBA), Uruguay.




\end{document}